\documentclass{ws-ijmpd}

\newcommand{\msun}{{M_{\odot}}}

\newcommand{\ls}{{\stackrel{\textstyle <}{_\sim}}}
\newcommand{\kFn}{{k_{F_n}}}
\newcommand{\fmmo}{{\rm fm}^{-1}}
\newcommand{\fmmt}{{\rm fm}^{-3}}
\newcommand{\mev}{{\rm MeV}}
\newcommand{\mevt}{{\rm MeV/fm}^3}
\newcommand{\eos}{equation of state~}

\newcommand{\okgr}{{\Omega_{\rm K}}}
\newcommand{\bag}{B^{1/4}}
\newcommand{\gcmt}{{\rm g/cm}^3}
\newcommand{\pkgr}{P_{\rm K}}
\newcommand{\msec}{\rm ms}
\newcommand{\hv}{HV}
\newcommand{\gthree}{${\rm G}_{300}^{\rm B180}$}
\newcommand{\nuk}{{\nu_{\rm K}}}

\begin{document}

\markboth{F.\ Weber et al.}  
{Strangeness in Neutron Stars}

\catchline{}{}{}{}{}

\title{Strangeness in Neutron Stars}

\author{FRIDOLIN WEBER\footnote{fweber@sciences.sdsu.edu}, ALEXANDER
HO\footnote{aho@rohan.sdsu.edu}, RODRIGO P.\
NEGREIROS\footnote{negreiro@sciences.sdsu.edu}, \\ PHILIP
ROSENFIELD\footnote{philrose@sciences.sdsu.edu}}

\address{Department of Physics, San Diego State University, 5500
Campanile Drive \\ San Diego, California 92182-1233,
USA}

\maketitle

\begin{history}
\received{Day Month Year}
\revised{Day Month Year}
\comby{Managing Editor}
\end{history}

\begin{abstract}
It is generally agreed on that the tremendous densities reached in the
centers of neutron stars provide a high-pressure environment in which
several intriguing particles processes may compete with each
other. These range from the generation of hyperons to quark
deconfinement to the formation of kaon condensates and H-matter. There
are theoretical suggestions of even more exotic processes inside
neutron stars, such as the formation of absolutely stable strange
quark matter. In the latter event, neutron stars would be largely
composed of strange quark matter possibly enveloped in a thin nuclear
crust.  This paper gives a brief overview of these striking physical
possibilities with an emphasis on the role played by strangeness in
neutron star matter, which constitutes compressed baryonic matter at
ultra-high baryon number density but low temperature which is no
accessible to relativistic heavy ion collision experiments.
\end{abstract}


\section{Introduction}

\begin{figure}[b] 
\centerline{\psfig{figure=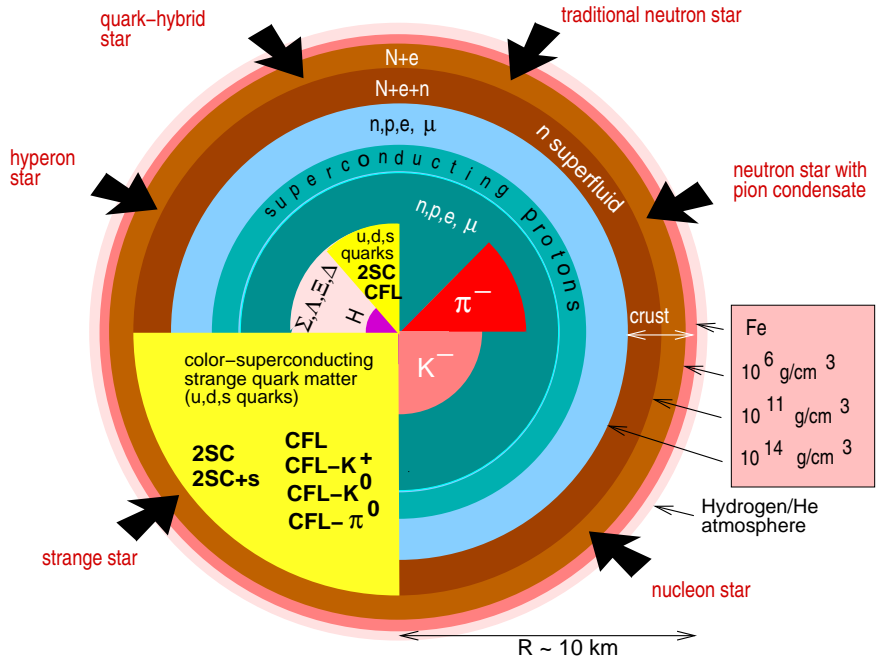,width=12.0cm,angle=0}}
\vspace*{8pt}
\caption[]{Competing structures and novel phases of subatomic matter
predicted by theory to make their appearances in the cores ($R\ls
8$~km) of neutron stars\cite{weber05:a}.}
\label{fig:cross}
\end{figure}

Neutron stars, spotted as pulsars by radio telescopes and x-ray
satellites, are among the most fascinating objects in the Universe.
They are more massive than our sun but are typically only about 10
kilometers across so that the matter in their centers is compressed to
densities that are up to an order of magnitude higher than the density
inside atomic nuclei\cite{glen97:book,weber99:book,sedrakian06:a}.  At
such extreme densities numerous subatomic particle processes are
expected to compete with each other and novel phases of matter--like
the quark-gluon plasma being sought at the most powerful terrestrial
particle colliders--could exist. Figure~\ref{fig:cross} summarizes the
situation graphically\cite{weber99:book,weber05:a}.
 The strangeness-carrying $s$ quark is likely to
play a key for the composition of neutron star matter, since several
potential building blocks of such matter contain $s$ quarks as one of
their constituents. Examples of which are the $\Lambda$, $\Sigma$ and
$\Xi$ hyperons, the $K^-$ meson, and the H-dibaryon.  First and
foremost, strangeness may also exist in the form of unconfined $s$
quarks, which could populate, in chemical equilibrium with $u$ and $d$
quarks and/or hadrons, extended regions inside neutron stars.  This
paper aims at giving a brief overview of the possible manifestations
of strangeness inside neutron stars.

\section{Proposed Particle Compositions}

\subsection{Hyperons}

Only in the most primitive conception, a neutron star is constituted
from neutrons.  At a more accurate representation, neutron stars will
contain neutrons, $n$, and a small number of protons, $p$, whose
charge is balanced by leptons, $e^-$, $\mu^-$.  At the densities that
exist in the interiors of neutron stars, the neutron chemical
potential, $\mu^n$, easily exceeds the mass of the $\Lambda$ so that
neutrons would be replaced with $\Lambda$ hyperons.  From the
threshold relation $\mu^n = \mu^\Lambda$ it follows that this would
happen for neutron Fermi momenta greater than $\kFn \sim 3 \, \fmmo$.
Such Fermi momenta correspond to densities of just $\sim 2 \rho_0$,
with $\rho_0 =0.16\, \fmmt$ the baryon number density of infinite
nuclear matter. Hence, in addition to nucleons and electrons, neutron
stars may be expected to contain considerable populations of
strangeness-carrying $\Lambda$ hyperons, possibly accompanied with
somewhat smaller populations of $\Sigma$ and $\Xi$
hyperons\cite{glen85:b}. The total hyperon population may be as large
as 20\% \cite{glen85:b}.

\subsection{Meson condensates and nucleon matter}

The condensation of negatively charged mesons in neutron star matter
is favored because such mesons would replace electrons with very high
Fermi momenta. Early estimates predicted the onset of a negatively
charged pion condensate at around $2 \rho_0$ (see, for instance, Ref.\
\refcite{baym78:a}). However, these estimates are very sensitive to
the strength of the effective nucleon particle-hole repulsion in the
isospin $T=1$, spin $S=1$ channel, described by the Landau
Fermi-liquid parameter $g'$, which tends to suppress the condensation
mechanism\cite{baeckman79:a}. Measurements in nuclei tend to indicate
that the repulsion is too strong to permit condensation in nuclear
matter\cite{barshay73:a,brown88:a}.  Nevertheless, some authors argue
to the contrary in the case of neutron star
matter\cite{tatsumi91:a,umeda92:a}. In the mid 1980s it was discovered
that the in-medium properties of $K^-$ mesons may be such that they
could condense in neutron star matter as
well\cite{kaplan86:a,brown87:a,lee95:a}.  Pion as well as kaon
condensates would have two important effects on neutron stars.
\begin{table}[htb]
\tbl{Properties of non-rotating neutron stars composed of nucleons and
hyperons (\hv), nucleons, hyperons, and normal quarks (\gthree), and
nucleons, hyperons, and color-superconducting quarks (CFL).}
{\begin{tabular}{@{}lccc@{}} \toprule
Stellar property & ~\hv~       & ~\gthree~  & ~CFL~         \\  \colrule
$\epsilon_{\rm c}~(\mevt)$ & 361.0\hphantom{00}  & 814.3\hphantom{00}  &2300.0\hphantom{00}  \\
$M$~($\msun$)                  & 1.39\hphantom{00}   & 1.40\hphantom{00}   & 1.36\hphantom{00}      \\
$R$~(km)                     & 14.1\hphantom{00}   & 12.2\hphantom{00}   & 9.0\hphantom{00}        \\
$Z$                          & 0.1889\hphantom{00} & 0.2322\hphantom{00} & 0.3356\hphantom{00}   \\
$g_{\rm {s,14}}~ ({\rm cm/s}^2)$ & 1.1086\hphantom{00} & 1.5447\hphantom{00} & 3.0146\hphantom{00}   \\
$BE~(\msun)$               & 0.0937\hphantom{00} & 0.1470\hphantom{00} & 0.1534\hphantom{00}   \\  \botrule
\end{tabular}\label{tab:1}}
\end{table}
Firstly, condensates soften the \eos above the critical density for
onset of condensation, which reduces the maximal possible neutron
mass. At the same time, however, the central stellar density
increases, because of the softening.  Secondly, meson condensates
would lead to neutrino luminosities which are considerably enhanced
over those of normal neutron star matter. This would speed up neutron
star cooling considerably\cite{thorsson94:a}.  The condensation of
$K^-$ mesons in neutron stars is initiated by the reaction
\begin{equation}
  e^- \rightarrow K^- + \nu \, , 
\label{eq:kaon.1}
\end{equation} 
which is shorthand for $p+e^- \rightarrow n + \nu$ and $n \rightarrow p
+ K^-$.  If this reaction becomes possible in a neutron star, it is
energetically advantageous for the star to replace the fermionic
electrons with the bosonic $K^-$ mesons. Whether or not this happens
depends on the behavior of the $K^-$ mass in neutron star matter.
Experiments which shed light on the properties of the $K^-$ in nuclear
matter have been performed with the Kaon Spectrometer (KaoS) and the
FOPI detector at the heavy-ion synchrotron SIS at
GSI\cite{barth97:a,senger01:a,sturm01:a,devismes02:a,fuchs06:a}.  An
analysis of the early $K^-$ kinetic energy spectra extracted from
Ni+Ni collisions \cite{barth97:a} showed that the attraction from
nuclear matter would bring the $K^-$ mass down to $m^*_{K^-}\simeq
200~\mev$ at $\rho\sim 3\, \rho_0$. For neutron-rich matter, the
relation
\begin{equation}
  m^*_{K^-}(\rho) \simeq m_{K^-} \left( 1 - \frac{1}{5} \,
  \frac{\rho}{\rho_0} \right)
\label{eq:meff02}
\end{equation} was established\cite{li97:a,li97:b,brown97:a}, with
$m_K = 495$~MeV the $K^-$ vacuum mass.  Values around $m^*_{K^-}\simeq
200~\mev$ lie in the vicinity of the electron chemical potential,
$\mu^e$, in neutron star matter\cite{weber99:book,glen85:b} so that
the threshold condition for the onset of $K^-$ condensation, $\mu^e =
m^*_K$, which follows from Eq.\ (\ref{eq:kaon.1}), could be fulfilled
in the centers of neutron stars. Equation (\ref{eq:kaon.1}) is
followed by
\begin{equation}
  n \rightarrow p + K^-  \, ,
\label{eq:npK.1}
\end{equation} with the neutrinos again leaving the star. 
By this conversion the nucleons in the cores of newly formed neutron
stars can become half neutrons and half protons, which lowers the
energy per baryon of the matter\cite{brown96:a}. The relatively
isospin symmetric composition achieved in this way resembles the one
of atomic nuclei, which are made up of roughly equal numbers of
neutrons and protons.  Neutron stars are therefore referred to in this
description as nucleon stars.  The maximal possible mass of nucleon 
stars, where Eq.\ (\ref{eq:npK.1}) has gone to completion, has been
calculated to be around $1.5\, \msun$
\cite{thorsson94:a}. Consequently, the collapsing core of a supernova
(e.g.\ 1987A), if heavier than this value, should go into a black hole
rather than forming a neutron
star\cite{li97:a,li97:b,brown94:a}. Another striking implication,
pointed out by Brown and Bethe, would be the existence of a large
number of low-mass black holes in our galaxy\cite{brown94:a}.

\subsection{H-dibaryons}

A novel particle that could make its appearance in the center of a neutron
star is Jaffe's H-dibaryon, a doubly strange six-quark composite with
spin and isospin zero, and baryon number two\cite{jaffe77:a}. In neutron star
matter, which may contain a significant fraction of $\Lambda$ hyperons, the
$\Lambda$'s could combine to form H-dibaryons which could give way to the
formation of H-matter at densities somewhere between $3\, \epsilon_0$ and $6\,
\epsilon_0$, depending on the in-medium properties of the H-dibaryon. H-matter
could thus exist in the cores of moderately dense neutron
stars\cite{glen98:a,tamagaki91:a,sakai97:a}. If formed, however, H-matter may
not remain dormant in the centers but, because of its instability against
compression, could trigger the conversion of neutron stars into hypothetical
strange stars\cite{sakai97:a,faessler97:a,faessler97:b}.

\subsection{Quark deconfinement}\label{ssec:deconf}

\begin{figure}[tb] 
\centerline{\psfig{file=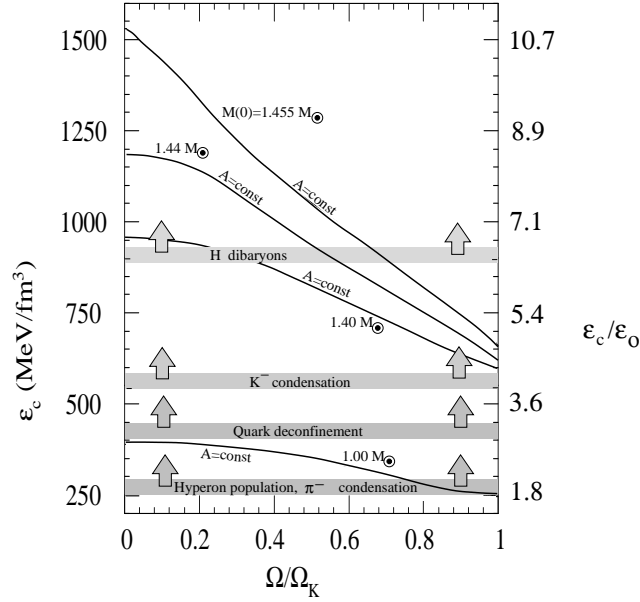,width=8.0cm,angle=-90}}
\vspace*{8pt}
\caption[]{Central density versus rotational frequency for several
sample neutron stars\cite{weber99:book}. The stars' baryon number, $A$,
is constant in each case.  Theory predicts that the interior stellar
density could become so great that the threshold densities of various
novel phases of superdense matter are reached. $\epsilon_0 =
140~\mevt$ denotes the density of nuclear matter, $\okgr$ is the
Kepler frequency, and $M(0)$ is the stars' mass at zero rotation.}
\label{fig:ec1445fig}
\end{figure}

It has been suggested already many decades
ago\cite{ivanenko65:a,fritzsch73:a,baym76:a,keister76:a,%
chap77:a,fech78:a,chap77:b} that neutrons, protons plus the heavier
constitutes ($\Sigma, ~\Lambda, ~\Xi, ~\Delta$) may melt under the
enormous pressure that exists in the cores of neutron stars, creating
a new state of matter know as quark matter.  At present one does not
know from experiment at what density the expected phase transition to
quark matter occurs, and one has no conclusive guide yet from lattice
QCD simulations either.  From simple geometrical considerations it
follows that nuclei begin to touch each other at densities around
$(4\pi r^3_N/3)^{-1} \simeq 0.24~\fmmt = 1.5\, \rho_0$ (less than
twice the density of nuclear matter!) for a characteristic nucleon
radius of $r_N\sim 1$~fm. This figure increases to $\sim 11 \, \rho_0$
for a nucleon radius of $r_N = 0.5$~fm. One may thus expect that the
nuclear boundaries of hadrons in the cores of neutron stars begin to
dissolve at densities somewhere between $\sim 2-10\, \rho_0$ and that
the quarks, originally confined into these hadrons, begin to populate
free states outside of them.  Depending on rotational frequency and
stellar mass, densities as large as two to three times $\rho_0$ are
easily surpassed in the cores of neutron stars of canonical mass, as
can be seen from Fig.\ \ref{fig:ec1445fig} and Tables \ref{tab:1} and
\ref{tab:2}\cite{weber06:a}, so that the neutrons and protons in the
centers of neutron stars may indeed have been broken up into their
constituent quarks by gravity\cite{weber99:topr}. More than that,
since the mass of the strange quark is only $m_s \sim 150$~MeV,
high-energetic up and down quarks will readily transform to strange
quarks at about the same density at which up and down quark
deconfinement sets\cite{weber05:a,glen91:pt,kettner94:b}, giving way
to the existence of three-flavor quark matter in the centers of
neutron stars\cite{glen97:book,weber99:book,weber99:topr,glen97:a}. We
also note that in contrast to relativistic heavy ion collisions, which
can only provide a fleeting glimpse of quark matter, neutron stars
would contain quark matter as a permanent component of matter in their
centers.
Whether or not quark deconfinement exists in static (non-rotating)
neutron stars makes only very little difference to their properties,
such as the range of possible masses and radii, which renders the
detection of quark matter in such objects extremely complicated. This
turns out to be
\begin{figure}[tb] 
\centerline{\psfig{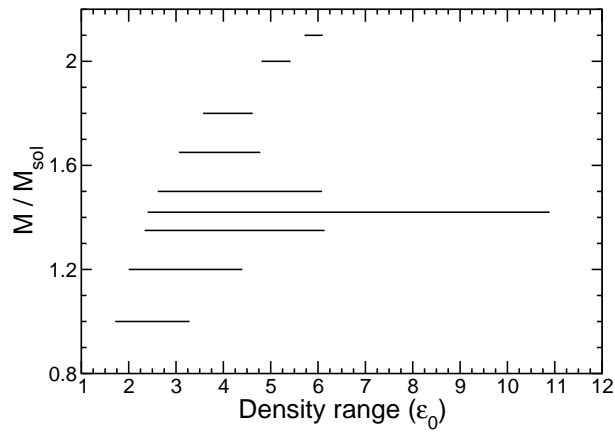}}
\vspace*{8pt}
\caption[]{Spread of central density (in units of the density of
nuclear matter, $\epsilon_0 = 140~\mevt$) of non-rotating neutron
stars for a broad collection of modern equations of
state\cite{weber99:book,weber05:a}. The very wide density range of
the $1.42\, \msun$ solar mass model has its origin in quark
deconfinement.}
\label{fig:density_ranges}
\end{figure} 
strikingly different for rotating neutron stars which develop quark
matter cores in the course of spin-down.  The reason being that as
such stars spin down, because of the emission of magnetic dipole
radiation and a wind of electron-positron pairs, they become more and
more compressed.  For some rotating neutron stars the mass and initial
\begin{table}[htb]  
\tbl{Same as Table~\ref{tab:1} but for neutron stars rotating at 
the Kepler frequency, $\nuk$.}
{\begin{tabular}{@{}lccc@{}} \toprule
Stellar property    & ~\hv~       & ~\gthree~  & ~CFL~     \\
                    & $\nuk$ = 850~Hz & $\nuk$ = 940~Hz  &$\nuk$ = 1400~Hz \\ 
\colrule
$\epsilon_{\rm c}~(\mevt)$ &280.0  &400.0 &1100.0  \\
$I~{\rm (km^{3})}$         & 223.6   & 217.1   & 131.8   \\
$M$~($\msun$)              & 1.39    & 1.40    & 1.41    \\
$R$~(km)                   & 17.1    & 16.0    & 12.6    \\
$Z_{\rm p}$                & 0.2374  & 0.2646  & 0.3618  \\
$Z_{\rm F}$                & $-$0.1788 & $-$0.1817 & $-$0.2184 \\
$Z_{\rm B}$                & 0.6046  & 0.6502  & 0.9190  \\
$g_{\rm {s,14}~ (cm/s^{2})}$ & 0.7278  & 0.8487  & 1.4493  \\
$T/W$                      & 0.0894  & 0.0941  & 0.0787  \\
$BE~(\msun)$               & 0.0524  & 0.1097  & 0.1203  \\
$V_{\rm eq}/c$             & 0.336   & 0.353   & 0.424   \\ \botrule
\end{tabular}\label{tab:2}}
\end{table}
rotational frequency may be just such that the central density rises
from below to above the critical density for the dissolution of
baryons into their quark constituents. This could effect the star's
moment of inertia dramatically\cite{glen97:a}. Depending on the rate
at which quark matter is produced, the moment of inertia can decrease
very anomalously, and could even introduce an era of stellar spin-up
(``backbending'') lasting for $\sim 10^8$ years\cite{glen97:a}. Since
the dipole age of millisecond pulsars is about $10^9$~years, one may
estimate that roughly about 10\% of the $\sim 30$ solitary millisecond
pulsars presently known could be in the quark transition epoch and
thus could be signaling the ongoing process of quark deconfinement.
Changes in the moment of inertia reflect themselves in the braking
index, $n$, of a rotating neutron star, as can be seen
from\cite{weber05:a,glen97:a,spyrou02:a}
\begin{equation}  
  n(\Omega) \equiv \frac{\Omega\, \ddot{\Omega} }{\dot{\Omega}^2} = 3
  - \frac{ I + 3 \, I' \, \Omega + I'' \, \Omega^2 } {I + I' \,
  \Omega} \rightarrow 3 - \frac{ 3 \, I' \, \Omega + I'' \, \Omega^2
  } {2\, I + I' \, \Omega}
\label{eq:index}
\end{equation}  
where dots and primes denote derivatives with respect to time and
$\Omega$, respectively.  The last relation in (\ref{eq:index})
constitutes the non-relativistic limit of the braking
index\cite{glen95:a}. It is obvious that these expressions reduce to
the canonical limit, $n=3$, if the moment of inertia is completely
independent of frequency. Evidently, this is not the case for rapidly
rotating neutron stars, and it fails for stars that experience
pronounced internal changes (as possibly driven by phase transitions)
which alter the moment of inertia significantly. In Ref.~
\refcite{glen95:a} it was shown that the changes in the moment of
inertia caused by the gradual transformation of hadronic matter into
quark matter may lead to $n(\Omega) \rightarrow \pm \infty$ at the
transition frequency where pure quark matter is produced.  Such
dramatic anomalies in $n(\Omega)$ are not known for conventional
neutron stars (see, however, Ref.\ \refcite{zdunik05:a}), because
their moments of inertia appear to vary smoothly with
$\Omega$.\cite{weber99:book}  The future astrophysical observation of a
strong anomaly in the braking behavior of a pulsar may thus indicate
that quark deconfinement is occurring at the pulsar's center.

Accreting x-ray neutron stars provide a very interesting contrast to
the spin-down of isolated neutron stars discussed just above. These
x-ray neutron stars are being spun up by the accretion of matter from
a lower-mass ($M \ls 0.4 \msun$), less-dense companion.  If the
critical deconfinement density falls within that of canonical pulsars,
quark matter will already exist in them but will be spun out of x-ray
stars as their frequency increases during
accretion\cite{glen00:b,glen01:a}.

\subsection{Color superconductivity}\label{sec:css}

There has been much recent progress in our understanding of quark
matter, culminating in the discovery that if quark matter exists it
will be a color superconductor\cite{rajagopal01:a,alford01:a}.  The
phase diagram of such matter is very complex. At asymptotic densities
the ground state of QCD with a vanishing strange quark mass is the
color-flavor locked (CFL) phase. This phase is electrically neutral in
bulk for a significant range of chemical potentials and strange quark
masses\cite{rajagopal01:b}. If the strange quark mass is heavy enough
to be ignored, then up and down quarks may pair in the two-flavor
superconducting (2SC) phase.  Other possible condensation patters are
the CFL--$K^0$ phase\cite{bedaque01:a} and the color-spin locked
(2SC+s) phase\cite{schaefer00:a}. Depending on the condensation
pattern, the magnitude of the supefluid gap energy ranges from several
keV's to about one hundred MeV. Color superconductivity has been shown
to have consequences for a number of astrophysical phenomena ranging
from neutron star cooling, to the arrival times of supernova
neutrinos, to the evolution of neutron star magnetic fields,
rotational (r-mode) instabilities, and glitches in rotation
frequencies of
pulsar\cite{rajagopal01:a,alford01:a,rajagopal00:a,alford00:a,alford00:b,%
blaschke99:a}.  Aside from neutron star properties, an additional test
of color superconductivity may be provided by upcoming cosmic ray
space experiments such as AMS~\cite{ams01:homepage} and
ECCO\cite{ecco01:homepage}. As shown in Ref.~\refcite{madsen01:a},
finite lumps of color-flavor locked strange quark matter, which should
be present in cosmic rays if strange matter is the ground state of the
strong interaction (see Sect.~\ref{ssec:ss}), turn out to be
significantly more stable than strangelets without color-flavor
locking for wide ranges of parameters. In addition, strangelets made
of CFL strange matter obey a charge-mass relation of $Z/A \propto
A^{-1/3}$, which differs significantly from the charge-mass relation
of strangelets made of ordinary strange quark matter. In the latter
case, $Z/A$ would be constant for small baryon numbers $A$ and $Z/A
\propto A^{-2/3}$ for large
$A$\cite{madsen01:a,aarhus91:proc,madsen98:b}. This difference may
allow an experimental test of CFL locking in strange quark
matter\cite{madsen01:a}.

\section{Absolutely Stable Strange Quark Matter}
\label{ssec:ss}

It is most intriguing that for strange quark matter made of more than
a few hundred up, down, and strange quarks, the energy of strange
quark matter may be well below the energy of nuclear matter, $E/A=
930$~MeV\cite{bodmer71:a,witten84:a,terazawa89:a}. A simple estimate
indicates that for strange quark matter $E/A = 4 B \pi^2/ \mu^3$, so
that bag constants of $B=57~\mevt$ (i.e.  $\bag=145$~MeV) and
$B=85~\mevt$ ($\bag=160$~MeV) would place the energy per baryon of
such matter at $E/A=829$~MeV and 915~MeV, respectively, which
correspond obviously to strange quark matter which is absolutely bound
with respect to nuclear
matter\cite{madsen88:a,madsen93:a,madsen97:bsky}. If this were indeed
the case, neutron star matter would be metastable with respect to
strange quark matter, and all neutron stars could in fact be strange
quark stars\cite{madsen88:a,madsen93:a,madsen97:bsky}. As briefly
described in Sect.\ \ref{sec:css}, strange quark matter is expected to
be a color superconductor which, at extremely high densities, should
be in the CFL phase.  This phase is rigorously electrically neutral
with no electrons required\cite{rajagopal01:b}. For sufficiently large
strange quark masses, however, the low density regime of strange quark
matter is rather expected to form a 2-flavor superconductor (2SC) in
which electrons are present\cite{rajagopal01:a,alford01:a}. The
presence of electrons causes the formation of an electric dipole layer
on the surface of strange matter, with huge electric fields on the
order of $10^{19}$~V/cm, which enables strange quark matter stars to
be enveloped in nuclear crusts made of ordinary atomic
matter\cite{kettner94:b,alcock86:a,alcock88:a,stejner05:a}. The
maximal possible density at the base of the crust (inner crust
density) is determined by neutron drip, which occurs at about $4\times
10^{11}~\gcmt$ or somewhat below\cite{stejner05:a}.

Since the nuclear crust surrounding a strange star would be bound to
the star by gravity rather than confinement, the mass-radius
relationship of a strange matter star with a nuclear would be
qualitatively similar to the one of purely gravitationally bound
neutron stars or white dwarfs. The fact that strange stars with crusts
tend to possess somewhat smaller radii than neutron stars leads to
smaller mass shedding (Kepler) periods $\pkgr$ for strange stars. This
is obvious from the classical mass shedding expression
$\pkgr=2\pi\sqrt{R^3/M}$ which carries over to the full general
relativistic case\cite{weber99:book}. It was found that due to the
smaller radii of strange stars the complete sequence of such objects
(and not just those close to the mass peak, as is the case for
neutron stars) can sustain extremely rapid rotation well below 1~ms
\cite{weber93:b}. In particular, strange stars with a canonical pulsar
mass of around $1.45\,\msun$ have Kepler periods in the range of $0.55
~\msec \ls \pkgr \ls 0.8 ~ \msec$, depending on the thickness of the
nuclear curst and the bag constant\cite{weber93:b,glen92:crust}. This
range is to be compared with $\pkgr\sim 1~\msec$ obtained for standard
neutron stars of the same mass computed for standard equations of
state. 

The fact that strange stars can carry nuclear crusts is key to
reconcile strange stars with superbursts and soft x-ray transient
phenomenology\cite{stejner06:a}.

\section{Effects of a Net Electric Charge Distributions on the 
Equation of State of Relativistic Stars}

As pointed out just above, the electric field that may exist on the
surface of a strange quark star would be as high as
$10^{19}$~V/cm\cite{usov04:a}.\footnote{Such ultra-high fields would
make the vacuum unstable in free space. The Pauli principle, however,
prevents this from happening on the surface of a strange star.}  The
energy density associated with such ultra-high fields begins to have
an influence one the curvature of space as determined by Einstein's
field equation\cite{malheiro04:a,bekenstein71:a,felice95:a},
\begin{eqnarray}
G_{\nu}{}^{\mu} = R_{\nu}{}^{\mu} - \frac{1}{2} g_{\nu}{}^{\mu} R =
\frac{8 \pi G}{c^4} T_{\nu}{}^{\mu} \ ,
\label{eq:einstein}
\end{eqnarray}
where $G_{\nu}{}^{\mu}$ is the Einstein tensor, $R_{\nu}{}^{\mu}$ the
Ricci tensor, $g_{\nu}{}^{\mu}$ the metric tensor, $R$ the Ricci
scalar, and $G_{\nu}{}^{\mu}$ the energy-momentum tensor. In the
following, we will discuss briefly how the equation of state of a
compact star is modified by the presence of an ultra-strong interior
electric field. The consequences for strange stars, where the electric
field is located on the surface, are explored in a paper that is
currently under preparation\cite{negreiros06:a}.
\begin{figure}[htb]
\centerline{\psfig{file=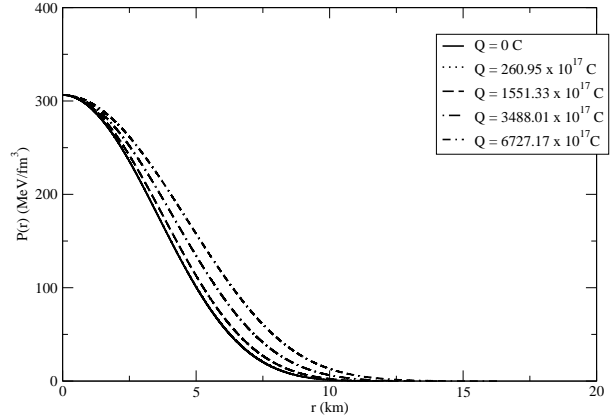,width=8.0cm}}
\vspace*{8pt}
\caption{Influence of electric charge on the pressure profiles of
compact stars. \label{pprof}}
\end{figure}
Our preliminary results indicate that, depending of the amount of
electric charge, the strange star structure and specifically the
mass-radius relationship might be drastically modified.  For the study
presented here we adopt a polytropic equation of state to model the
compact star\cite{malheiro04:a}.  First we will briefly present the
structure equations for a electrically charged star and then show how
the polytropic equation of state is modified by the presence of
charge.

\subsection{Energy-momentum tensor} 

We want to maintain spherical symmetry of the star. Thus the natural
choice for the metric is 
\begin{equation}
ds^2=e^{\nu(r)}c^{2}dt^{2}-e^{\lambda(r)}dr^{2}-r^{2}(d\theta^{2}
+\sin^{2}\theta d\phi^{2}) \, . \label{metr}
\end{equation}
The energy-momentum tensor of the star consists of two terms, the
standard term which describes the star's matter as a perfect fluid and
the electromagnetic term,
\begin{equation}
T_{\nu}^{}{\mu} = (p +\rho c^2)u_{\nu} u^{\mu} + p
\delta_{\nu}{}^{\mu} + \frac{1}{4\pi} \left[ F^{\mu l} F_{\nu l}
+\frac{1}{4 \pi} \delta_{\nu} ^{\mu} F_{kl} F^{kl} \right] \, ,
\end{equation}
where the components $F^{\nu \mu}$ satisfy the covariant Maxwell
equations, 
\begin{equation}
[(-g)^{1/2} F^{\nu \mu}]_{, \mu} = 4\pi J^{\nu} (-g)^{1/2} \label{ecem},
\end{equation}
where $J^\nu$ is the four-current. Imposing that in last equation the
only non-vanishing term is the radial component, one obtains for the
electric field
\begin{equation}
F^{01}(r)= E(r) = e^{-(\nu + \lambda)/2} \, r^{-2} \int_{0}^{r} 4\pi
j^{0} e^{(\nu + \lambda )/2} dr' \, . \label{comp01}
\end{equation}
From last equation we can define the charge of the system as
\begin{equation}
Q(r) = \int_{0}^{r} 4\pi j^{0} r'^{2} e^{(\nu +\lambda)/2} dr'. \label{Q}
\end{equation}
The electric field is then given by
\begin{equation}
E(r) = e^{-(\nu +\mu)/2} \, r^{-2} \, Q(r) \, .
\end{equation}
With the aid of these relations the energy-momentum tensor takes the
following form
\begin{equation}
T_{\nu} ^{\mu} =\left( \begin{array}{cccc}
-\left( \epsilon + \frac{Q^{2} (r)}{8\pi r^4} \right)  & 0 & 0 & 0 \\
0 & p - \frac{Q^{2} (r)}{8\pi r^4 } & 0 & 0 \\
0 & 0 & p + \frac{Q^{2} (r)}{8\pi r^4 }  & 0 \\
0 & 0 & 0 & p  +\frac{Q^{2} (r)}{8\pi r^4 }
\end{array} \right) \, . \label{TEMch}
\end{equation}
Substituting this relation into Einstein's field equation
(\ref{eq:einstein}) one arrives at
\begin{eqnarray}
e^{-\lambda}\left(
-\frac{1}{r^{2}}+\frac{1}{r}\frac{d\lambda}{dr}\right)
+\frac{1}{r^{2}}=\frac{8\pi G}{c^4} \left( p - \frac{Q^{2}(r)}{8\pi
  r^4} \right) \, , \label{fe1q} \\
e^{-\lambda}\left(\frac{1}{r}\frac{d\nu}{dr}+\frac{1}{r^{2}}\right)
-\frac{1}{r^{2}}= - \frac{8 \pi G}{c^4} \left( \epsilon +
\frac{Q^{2}(r)}{8\pi r^4} \right) \, . \label{fe2q}
\end{eqnarray}
The solution for the metric function $\lambda(r)$ is given by
\begin{equation}
e^{-\lambda} = 1 - \frac{Gm(r)}{rc^2} +\frac{GQ^2}{r^2 c^4} \,
. \label{metd}
\end{equation}
The first two terms on the right-hand-side of this relation correspond
to electrically neutral stars, while the third term originates from
the net electric charge distribution inside the star. Using Eq.\
(\ref{metd}) together with Eqs.\ (\ref{fe1q}) and (\ref{fe2q}) we
arrive at an equation for the mass of the star within a spherical 
shell of radius, $m(r)$,
\begin{equation}
\frac{dm(r)}{dr} = \frac{4\pi r^2}{c^{2}} \epsilon +\frac{Q(r)}{c^2
r}\frac{dQ(r)}{dr} \, . \label{dmel}
\end{equation}
The first term on the right-hand-side is the standard result for the
gravitational mass of electrically uncharged stars, while the second
term accounts for the mass change that originates from the electric
field.

\subsection{Tolman-Oppenheimer-Volkoff equation}

\begin{figure}[htb]
\centerline{\psfig{file=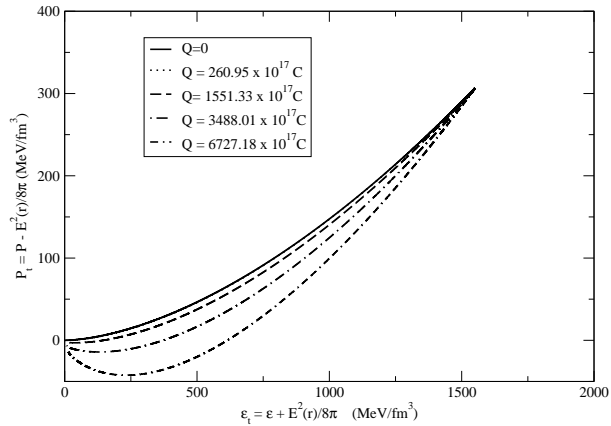,width=8.0cm}}
\vspace*{8pt}
\caption{Influence of electric charge distribution on the equation
of state of a compact star. \label{EOS}}
\end{figure}

The equation of general relativistic hydrostatic equilibrium, known as
the Tolman-Oppenheimer-Vokoff (TOV) equation, are obtained after
imposing $T_{\nu}{}^{\mu}{}_{;\mu} =0$. This leads to the
\begin{equation}
\frac{dp}{dr} = - \frac{2G\left( m(r) +\frac{4\pi r^3}{c^2} \left( p -
\frac{Q^{2} (r)}{4\pi r^{4} c^{2}} \right) \right)}{c^{2} r^{2} \left(
1 - \frac{2Gm(r)}{c^{2} r} + \frac{G Q^{2}(r)}{r^{2} c^{4}} \right)}
(p +\epsilon) +\frac{Q(r)}{4 \pi r^4}\frac{dQ(r)}{dr} \, . \label{TOVca}
\end{equation}
This completes the derivation of the structure equations of a
electrically charged compact star. Before we go ahead and solve the
TOV equation, we need to make an assumption for the charge
distribution inside the star. Only then the problem is fully
defined. Here we will follow the approach of Ref.~
\refcite{malheiro04:a,ray03:a} and assume that the charge distribution
is proportional to the energy density,
\begin{equation}
j^0(r) = f \times \epsilon \, ,
\end{equation}
where $f$ is a constant which essentially controls the amount of net
electric charge carried by the star. Adopting as initial and boundary
conditions of the problem the following values,
\begin{eqnarray}
\epsilon (0) = 1550 \quad \text{MeV}/\text{fm}^3 \quad \quad Q(0)
= 0, \\ m(0) = 0, \quad \quad \lambda(0) =0, \\ \nu(0) =0, \quad \quad
p(R) =0 \, ,
\end{eqnarray}
the properties of five different charged sample stars are compiled in
Table \ref{res}. As one might expect, both  mass and radius of a
compact star increase with net electric stellar charge.
\begin{table}[pb]
\tbl{Results} {\begin{tabular}{@{}lclr@{}} \toprule $f$ & $M~ (M_\odot) 
$& Radius (km) & Charge ($ \times 10^{17}$ C) \\
\colrule
0.0    & 1.428 & 11.85  & 0   \\
0.0001 & 1.439 & 11.879 & 260.95  \\
0.0005 & 1.742 & 12.559 & 1551.33 \\
0.0008 & 2.548 & 14.026 & 3488.01  \\ 
0.001  & 4.156 & 16.364 & 6727.17  \\ \botrule
\end{tabular} \label{res}}
\end{table}
These features have their origin in the repulsive pressure force
introduced into the system by the electric field. Figure \ref{pprof}
shows that in stellar configurations with higher electric charge, the
pressure drops at a slower rate, leading to bigger and therefore more
massive stars. In Fig.\ \ref{EOS} we show the effective (baryonic and
electric) equation of state of an electrically charged compact
star. 
The solid line is for electrically neutral (uncharged) stars computed
for the polytropic equation of state. The other curves are for stars
with successively increasing charges. Charges on this order of
magnitude may exist in the surface region of strange stars which, as
already mentioned above, is currently under
investigation\cite{negreiros06:a}.

\section{Summary}

It is often stressed that there has never been a more exciting time in
the overlapping areas of nuclear physics, particle physics and
relativistic astrophysics than today.  This comes at a time where new
orbiting observatories such as the Hubble Space Telescope (HST), Rossi
X-ray Timing Explorer, Chandra X-ray satellite, and the X-ray Multi
Mirror Mission (XMM) have extended our vision tremendously, allowing
us to observe compact star phenomena with an unprecedented clarity and
angular resolution that previously were only imagined.  On the Earth,
radio telescopes (Arecibo, Green Bank, Parkes, VLA) and instruments
using adaptive optics and other revolutionary techniques have exceeded
previous expectations of what can be accomplished from the
ground. Finally, the gravitational wave detectors LIGO, LISA, and
VIRGO are opening up a window for the detection of gravitational waves
emitted from compact stellar objects such as neutron stars and black
holes. This unprecedented situation is providing us with key
information on compact stars.  As discussed in this paper, a key role
in compact star physics is played by strangeness. It alters the
masses, radii, cooling behavior, and surface composition of neutron
stars, and may even give rise to new classes (strange stars, strange
dwarfs) of compact stars.  Other important observables may be the
spin evolution of isolated neutron stars and neutron stars in low-mass
x-ray binaries. All told, these observables are key in exploring the
phase diagram of dense nuclear matter at high baryon number density
but low temperature, which is not accessible to relativistic heavy ion
collision experiments.

\section*{Acknowledgments}

\section{References}

The research of F.\ Weber is supported by the National Science
Foundation under Grant PHY-0457329, and by the Research Corporation.

\end{document}